# Secure End-to-End Communications with Lightweight Cryptographic Algorithm

Augustine Ukpebor, James Addy, Kamal Ali and Ali Abu-El Humos

Jackson State University, Jackson, MS 39217, USA
augustine.ukpebor@students.jsums.edu, james.c.addy@students.jsums.edu,
kamal.ali@jsums.edu, and ali.abu-el_humos@jsums.edu

**Abstract:** The field of lightweight cryptography has been gaining popularity as traditional cryptographic techniques are challenging to implement in resource-limited environments. This research paper presents an approach to utilizing the ESP32 microcontroller as a hardware platform to implement a lightweight cryptographic algorithm. Our approach employs KATAN32, the smallest block cipher of the KATAN family, with an 80-bit key and 32-bit blocks. The algorithm requires less computational power as it employs an 80 unsigned 64-bit integer key for encrypting and decrypting data. During encryption, a data array is passed into the encryption function with a key, which is then used to fill a buffer with an encrypted array. Similarly, the decryption function utilizes a buffer to fill an array of original data in the form of 32 unsigned 64-bit integers. This study also investigates the optimal implementation of cryptography block ciphers, benchmarking performance against various metrics, including memory requirements (RAM), throughput, power consumption, and security. Our implementation demonstrates that data can be securely transmitted end-to-end with good throughput and low power consumption.

## I. INTRODUCTION

Embedded Internet of Things (IOT) and resource constrained devices such as network sensors and RFID, are in huge demand today. Unfortunately, there are growing security and privacy concerns regarding these devices leaving them vulnerable to attacks by stronger class of adversaries. These concerns have prompted security researchers in the past decades to conduct series of research on lightweight cryptography to address the security issues.

A lightweight cipher is a cryptographic algorithm targeted for low-resource device, optimized for minimal area and/or memory overhead, low-power design, and adequate security level [1, 2]. Compared with conventional ciphers, a lightweight cipher has smaller block size, smaller key size, and implements simpler operations and scheduling techniques [3]. The main purpose of lightweight algorithms is to reduce the size of the main parameters of the algorithm, which includes block size, key size, fit to constrained devices, maximum throughput with minimum energy and memory consumptions [4]. Table 1 below provides various cryptography, examples, and applications.

**Table 1** Cryptography ciphers

| Cryptography | Examples | Applications |
|---|---|---|
| Conventional cryptography | DES (Data Encryption Standard), 3DES | Systems, Networks |
| Lightweight cryptography | PRESENT, KATAN | IOT devices, RFID, Embedded systems |

Conventional cryptography, known also as secret-key cryptography or symmetric-key encryption which converts plain text into cipher text could be block or stream ciphers. Block cipher uses either 64 bits or more and converts the plain text by taking its block at a time whereas stream cipher uses one byte and converts the plain text by taking one byte of the plain text at a time.

Two variants of lightweight cryptography implementation exist namely: hardware and software. In hardware implementations, chip size and/or energy consumption are the important measures to evaluate the lightweight properties. In software implementations, the smaller code and/or RAM size are preferable for the lightweight applications [5]. This research focuses on the hardware deployment using ESP32 microcontroller to implement lightweight cryptographic algorithm, KATAN. We use the smallest cipher of this family, KATAN32. The KATAN32 has a gate equivalent (GE) of 462GE while achieving encryption speed of 12.5 KBit/sec at 100 KHz [6].

In the year 2018, NIST issued submission requirements and evaluation criteria for the lightweight cryptography standardization process. The evaluation of candidates is based on specific criteria such as security, cost metrics (area, memory, energy consumption), performance, third party analysis and suitability for hardware and software implementation. The assessment is in the conclusive stage and is expected to last approximately 12 months [22].

This paper is structured so that section II discusses and compares some of the lightweight ciphers. Section III describes the system architecture in detail. The results are analyzed and evaluated in section IV. Finally, section V summarizes the paper and presents future directions.

## II. Lightweight Cryptography

In embedded systems security, lightweight terminology is used to essentially signify that an algorithm is suitable for use on some constrained environments. Lightweight cryptography is a cryptographic algorithm tailored to develop fast and efficient cryptographic mechanisms for implementation in constrained environments including RFID tags, sensors, contactless smart cards, health-care devices and so on [7]. Lightweight cryptography is desirable because of its high performance and security compare to the conventional cryptography. It is a resourceful tool for efficient implementations of cryptographic primitives. The motivation of lightweight cryptography is to use less memory, less computing resource, and less power supply to provide security solution that can work over resource-limited devices. The lightweight cryptography is, as expected, simpler and faster compared to conventional cryptography [8]. The goal of lightweight cryptography is to enable a diverse range of modern applications, such as smart meters, vehicle security systems, wireless patient monitoring systems, Intelligent Transport Systems (ITS) and the Internet of Things (IoT). The disadvantage of lightweight cryptography is that it is less secured [9]. There are many parameters that contribute to the efficiency of a given lightweight design, with area, power, throughput and energy being the foremost among them [10].

Lightweight cryptography is an encryption method that features a small footprint and/or low computational complexity [11]. Due to the harsh cost constraints and a very strong attacker model—especially noteworthy is the possibility of physical attacks—there is an increasing need for lightweight security solutions that are tailored to the ubiquitous computing paradigm [12].

### A. Lightweight Ciphers Comparison

Table 2 below is a comparison of some of the lightweight ciphers [6, 14].

**Table 2** Comparison of various lightweight ciphers

| Cipher Name | Cipher Type | Key Size | Block Size | Throughput (Kbps) | Area (GE) | Logic Process |
|---|---|---|---|---|---|---|
| KATAN32 [6] | **Block cipher** | 80 | 32 | 12.5 | 802 | 0.13 $\mu$m |
| AES-128 [13] | | 128 | 128 | 12.5 | 3400 | 0.35 $\mu$m |
| PRESENT [14] | | 80 | 64 | 200 | 1570 | 0.18 $\mu$m |
| HIGHT [15] | | 128 | 64 | 6400 | 3048 | 0.25 $\mu$m |
| | **Stream Cipher** | | | | | |
| Grain [16] | | 1 | 1 | 100 | 2599 | 0.13 $\mu$m |
| Trivium [17] | | 1 | 1 | 100 | 1294 | 0.13 $\mu$m |

### B. Benefits of Lightweight Block Ciphers Over Conventional Cryptography

To achieve the benefits of lightweight block ciphers over conventional cryptography block ciphers, the following design specifications informed the choice of KATAN for this paper:

1. **Size of device RAM, ROM, Circuit:** Constrained devices like ESP32 have limited memory and implementing conventional cryptography on them becomes almost practically impossible.
2. **Power consumption:** Some energy harvesting and battery-driven devices consume a lot of power and IOT device such as ESP32 is not exempt.
3. **Throughput:** A high throughput is necessary for devices with large data transmissions in the case of this research where volumes of live data are transmitted to the Internet. Using the conventional cryptography adds significant overheads which could invariably slow down the connection between the sensor systems and the cloud.
4. **Small key size**: KATAN lightweight block cipher uses 80 bits key sizes for performance and efficiency.
5. **Small block size:** Lightweight block ciphers use smaller block size, for instance KATAN is composed of three block ciphers, with 32, 48, or 64-bit block sizes whereas the traditional cryptography block cipher like AES uses 128-bit. It should also be noted that using small block sizes reduces limit on the maximum number of plaintext blocks to be encrypted. For example, outputs of a 64-bit block cipher can be distinguished from a random sequence using around 232 blocks for some of the approved modes of operations. Depending on the algorithm, this may lead to attacks such as plaintext recovery or key recovery or with non-negligible probabilities [18].
6. **Simple key schedule:** Complex key schedules increase the memory, latency, and the power consumption of implementations; therefore, most of the lightweight block ciphers use simple key schedules that can generate sub-keys on the fly [18]. This may enable attacks using related keys, weak keys, known keys or even chosen keys. Using a secure key derivation function (KDF) can prevent some of these attacks (for examples, see [19]).
7. **Simpler rounds:** The components and operations used in lightweight block ciphers are typically simpler than those of conventional block ciphers [18]. In lightweight designs using S-boxes, 4-bit S-boxes are preferred over 8-bit S-boxes. This reduction in size results in significant area savings. For example, the 4-bit S-box used in PRESENT required 28 GEs, whereas the AES S-box required 395 GEs in [20].

**III. SYSTEM DESCRIPTION**

The system architecture comprises of three sections: Microcontroller Unit (MCU) – ESP32, the perimeter devices, and the Internet web server. The security system is shown in figure 1 below which provides the general overview of the physical deployment of the system. The ESP32 is a feature rich MCU with integrated Wi-Fi with low power consumption. The user takes full control of the microcontroller through USB cable connection to carry out the necessary configurations on the unit. The microcontroller searches for the available WiFi and establish secure connection to the on-premises WiFi equipment (wireless router). Prior to the WiFi connectivity, the user configures the security modes in the authentication stage and credentials that the MCU uses to connect to the available WiFi. To enhance secure connection to WiFi, WPA2 Enterprise is chosen, the WPA2 Enterprise uses IEEE 802.1X standard which offers enterprise-grade authentication.

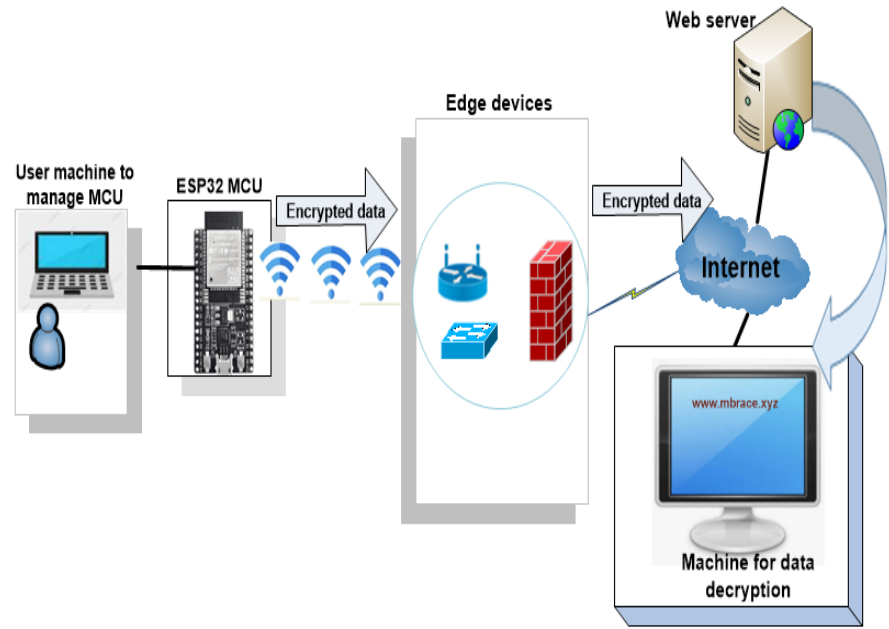

**Fig. 1** System architecture

Once a secure connection is established and data transmission is initiated, the KATAN32 algorithm encrypts the data and securely transmits the encrypted and encoded data through the perimeter devices to the web server on the Internet. All data received by the web server remain encrypted at rest. To decrypt the data, a user can use any machine to download the encrypted data from the server, www.mbrace.xyz and create an executable file from KATAN32 to decrypt the data. The techniques of the encryption and decryption algorithms are discussed in detail in the results section. Again, the block diagram (figure 2) analyzes the flow process starting from when plain text is encrypted to when cipher text is decrypted.

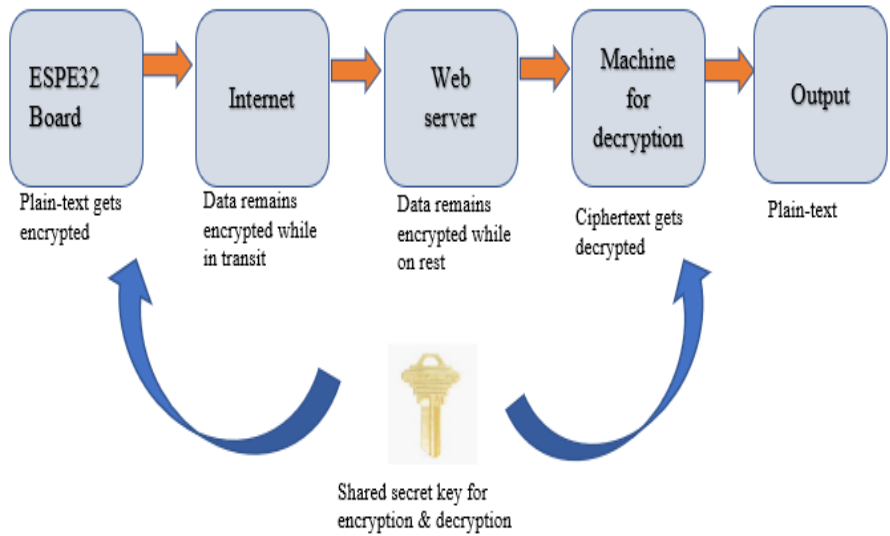

**Fig. 2** Symmetric-key flow process

## IV. EVALUATION AND RESULTS

### A. Evaluation

This implementation achieves great speed with low power consumption while demonstrating symmetric-key cryptography using KATAN32 lightweight block cipher. This same code is used to encrypt and decrypt the sample data sent to the server. The source code is written in C programming language by [21].

The implementation's functions receive portions of data in the form of an array of 32 unsigned 64-bit integers. This means that since the data is read as bytes, the bytes must be organized into 64-bit values. This is done by converting every 8 bytes of data into a 64-bit integer. Additional functions were written to convert bytes of data into an array of unsigned 64-bit integers as well as to convert back to bytes from the 64-bit integers. Since the functions receive 32 units of 64-bit integers, 256 bytes of data can be converted and passed into the encryption function at a time. Larger amount of data can simply be broken down into 256-byte portions. If a chunk of data is less than 256 bytes, the remaining elements are padded with zeros.

This algorithm uses a key of 80 unsigned 64-bit integers to encrypt and decrypt data. Once the data is converted into the array, it can be passed into the encryption function with the key where a buffer is filled with an encrypted array. When decrypting, the same key is used on the encrypted data and the decryption function will fill a buffer with an array of the original data in the form of 32 unsigned 64-bit integers.

The data is sampled in binary form and needs to be sent to a remote server. Since the data is received in binary and encrypted, it is not guaranteed to be safely formatted for HTTP requests. Therefore, once the data is encrypted, it is encoded using Base64 encoding. The encoded data can then be transmitted to the remote server where the data is Base64 decoded and stored in a binary file. The encrypted file can then be downloaded and decrypted using an executable written in C. The executable utilizes the KATAN32 decryption function and then locally creates and writes to a file containing the decrypted data.

The code for testing the encryption was written for the Esp32 and Esp8266 development boards using the Arduino Sketch IDE. These boards have a Wi-Fi module that allows for connection and communication with the remote file server. The remote server is provided using a web hosting service. The test code takes example data, encrypts, and Base64 encodes it. The encrypted and encoded data is wrapped in JSON format and sent to the server as an HTTP POST request. The server has PHP code that reads the data, decodes it, and writes it to a binary file. The data is still encrypted in this file. The file can be downloaded onto a local computer and then decrypted using the executable written in C. Both the code for the board that encrypts the data and the code for the executable that decrypts the data are hardcoded with identical keys. The test proved successful in demonstrating how sample data can be securely transferred using the KATAN32 implementation.

### B. Optimal Implementation of Cryptography Block Ciphers

This section presents implementation of lightweight cryptography primitive - KATAN and compares with the PRESENT and conventional cryptography, AES. We benchmark the performance against preferable metrics: memory requirements (RAM), throughput, power consumption, and security. A low-cost microcontroller, ESP32 was used for the study with simulation setup in table 3. The ESP32 development board (38 pins) is a microcontroller equipped with ESP-WROOM-32 module. The board is fitted with 2.4 GHz dual-mode WiFi, Bluetooth chips as well as the 40nm low-power technology. It has flash memory (program space) that stores the Arduino sketch. Static random-access memory (SRAM) is where the sketch creates and manipulates variables when it runs [23].

**Table 3**: Simulation and synthesis setup

| # | Description | Specification |
|---|---|---|
| 1 | Device | Microcontroller Esp32 |
| 2 | Frequency | 240 MHz |
| 3 | Memory | 520 KiB internal SRAM |
| 4 | Flash | 4MB external flash |
| 5 | ROM | 448 KB of ROM |

### C. KATAN

Generally, KATAN is a hardware-oriented block cipher with the same basic encryption algorithm but with different key scheduling and block size. KATAN is composed of three block ciphers, with 32, 48, or 64-bit block size with 80-bit key size [27] for performance and efficiency. It is three times faster for negligible area overhead, hence one of the smallest known ciphers [26]. The interesting feature about this cipher is that it uses fixed key size to further simplify the hardware during encryption and decryption. Several surveys and studies highlighted the superior performance of KATAN ciphers [24].

### D. PRESENT

It is called Ultra Lightweight Cryptographic Algorithm because it gives an ultra-lightweight solution and sufficient security when compared to other algorithms with key size of 80/128 bits on constrained devices [28]. It has Substitution-Permutation networks (SPN) structure and works on 64-bit block-size. It is hardware efficient, and has 31rounds which consist of XOR, and a non-linear layer 4-bit S-box which is applied 16 times in parallel [25].

### E. Advanced Encryption Standard (AES)

AES is a symmetric block cipher that has a data block of fixed 128 bits with variable key size of 128, 192 or 256 bits. It is based on Substitution-Permutation networks (SPN). AES is an iterative algorithm, and each iteration is called a round, depending on the key length 128/192/256, the number of total rounds chosen are 10/12/14 respectively [25]; it works on 4*4 matrix, where there are four processes in each round: add round key, byte substitution, shift row and mix columns.

### F. Results and Performance Analysis

Table 4 summarizes the implementations of the block ciphers in this study. Four parameters were used to evaluate and compare the ciphers: memory, throughput, power consumption, and security.

**Table 4** Analysis of Experimental Results of KATAN 32 Block Cipher

| # | Block Cipher | Block size | Key Size | Code size (bytes) | Throughput (Kbps) | Power consumption (mW) |
|---|---|---|---|---|---|---|
| 1 | KATAN | 32 | 80 | 3,104 | 11.12 | 105 |
| 2 | PRESENT [6, 30] | 64 | 80 | 157,472 | 11.4 | 114 |

| 3 | AES [29, 30] | 128 | 128 | 438,050 | 12.4 | 115 |

The first metric considered to evaluate the algorithms is memory consumption. The code when loaded on the ESP32 flash memory was relatively small compared to that of PRESENT and AES, see figure 3. Comparing the three ciphers, the PRESENT consumes the highest memory. Although KATAN32 is regarded as one of the slowest ciphers, our experiment was comparatively fast perhaps because of the speed of the processor (240MHZ); the throughput of KATAN32 was almost the same as that of PRESENT and AES as shown in the experimental results, see figure 4.

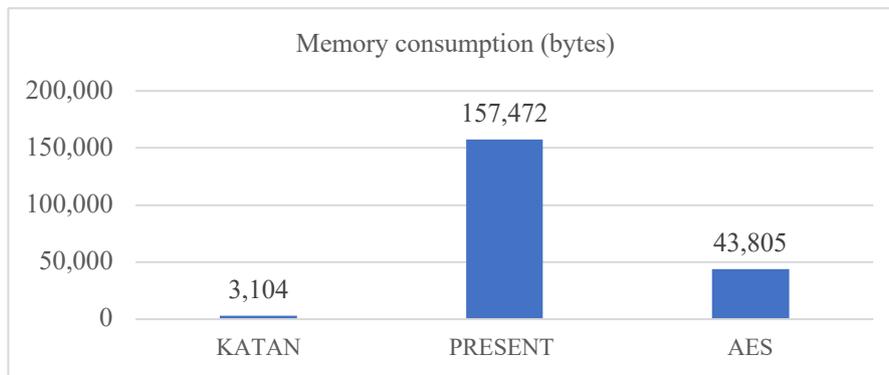

**Fig. 3** Memory consumption on ESP32

The throughput was measured by sending data from the source to the destination. Using the Arduino timestamp, the start and the end times were recorded to compute the execution time. Finally, the throughput was calculated using the formula as follows:

Throughput = Data (in bytes) / (end time - start time).

The more data executed, the higher the precision of the measurement as elucidated in table 5.

The data transmitting rate was slow due to slow internet connection. Given that 250 bytes were transmitted per time, to compute data rate, we have: $125*10^6 / 250 = 500,000$ sends per second. In other word, data will be sent every $1 / 500,000 = 0.000002$ second = 2 microseconds in order to achieve a transmission rate of 1 Gbps. This was not achieved because of the limited broadband see figure 5. This figure shows the rate of data transmitted over a transmission channel within a specified unit of time, bits/s.

The power consumption of KATN32 was relatively lower than other ciphers as shown in figure 6. The significant observation among the three ciphers is that AES ranks highest in terms of power consumption. It can be inferred that the higher the throughput, the more power consumption as seen in the AES cipher.

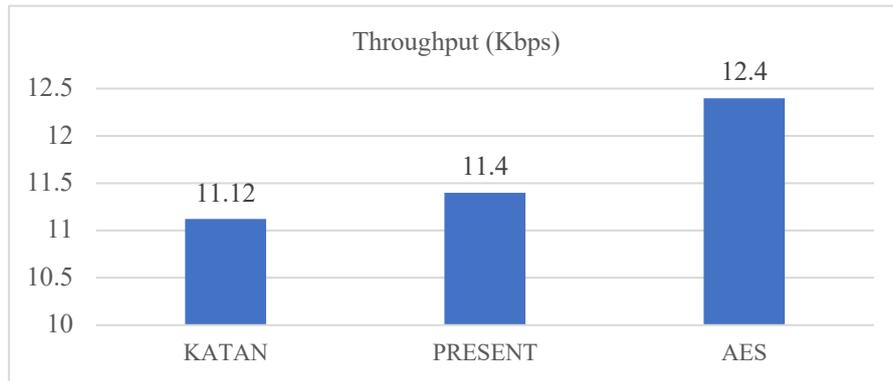

**Fig. 4** Throughput

**Table 5:** Throughput of KATAN32 Block Cipher

| Bytes | Bits | Milli-seconds | Seconds | Throughput (b/s) | Through-put (Kb/s) |
|---|---|---|---|---|---|
| 250 | 2000 | 105 | 0.105 | 19047.62 | 19.05 |
| 250 | 2000 | 210 | 0.21 | 9523.81 | 9.52 |
| 250 | 2000 | 176 | 0.176 | 11363.64 | 11.36 |
| 250 | 2000 | 286 | 0.286 | 6993.01 | 6.99 |
| 250 | 2000 | 150 | 0.15 | 13333.33 | 13.33 |
| 250 | 2000 | 182 | 0.182 | 10989.01 | 10.99 |
| 250 | 2000 | 208 | 0.208 | 9615.38 | 9.62 |
| 250 | 2000 | 208 | 0.208 | 9615.38 | 9.62 |
| 250 | 2000 | 208 | 0.208 | 9615.38 | 9.62 |
| Average throughput | | | | | 11.12 |

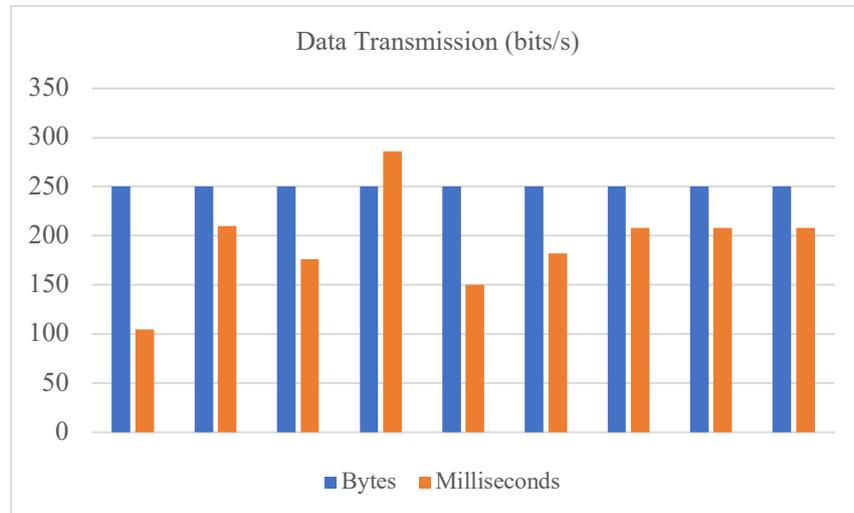

**Fig. 5** Data transmission

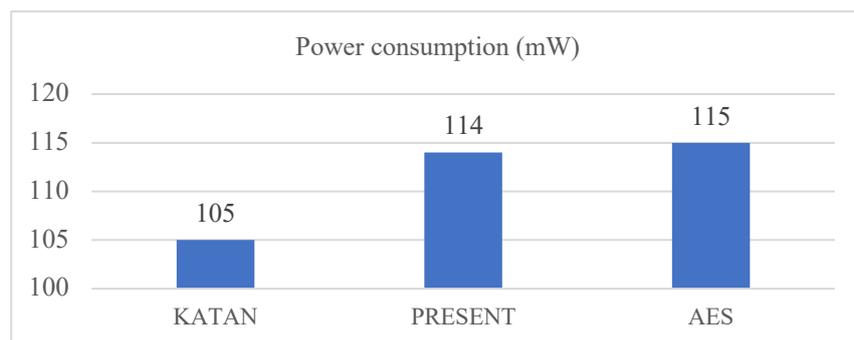

**Fig. 6** Power consumption

Another metric that was assessed is the security of KATAN32 algorithm. The strength of KATAN32 cipher was tested with a third-party application, Fiddler, by intercepting encrypted web traffics. Fiddler is capable of capturing HTTP/HTTPS traffics from any browser; conducts web penetration testing by decrypting the contents of the encrypted web traffics. It is a handing and powerful tool for manipulating sessions and requests. By default, Fiddler does not capture and decrypt secure HTTPS traffic. To log data sent through HTTPS, we enabled HTTPS traffic decryption.

We initiated encrypted traffic from the ESP32 development board to the Internet and used the Fiddler to capture HTTPS traffics from the browser. As expected, the captured web traffics remained encrypted, see figure 7 below.

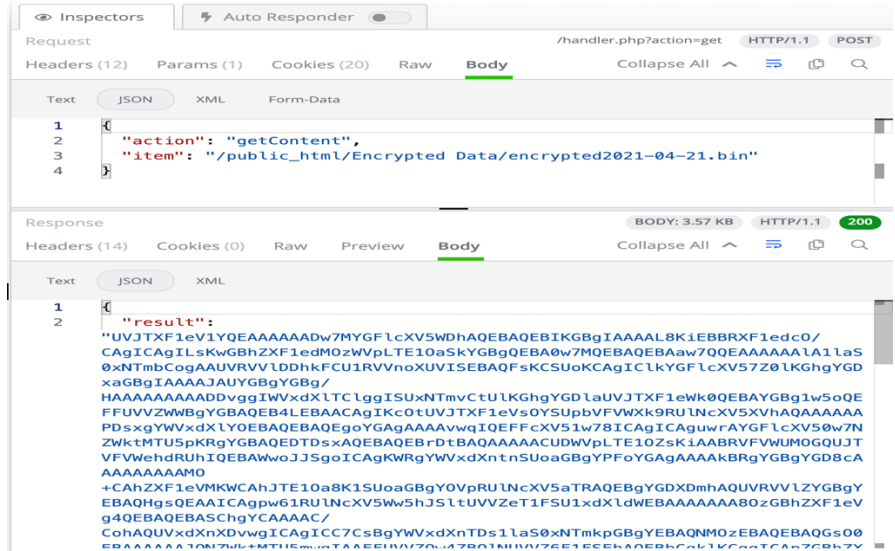

**Fig. 7** Binary data showing the time data arrived on the server

## V. CONCLUSION

This paper investigates the feasibility of ultra-lightweight cryptography for real-time data transmission over the internet. Specifically, we implement the KATAN32 block cipher and compare its performance with other ciphers, namely PRESENT and AES. Our approach is tailored towards resource-constrained environments, ensuring end-to-end security from the point of deployment to data storage. The key achievements of our research include achieving fast data transfer speeds and ensuring a secure channel for data transmission. Moving forward, we suggest exploring the development of an algorithm to automatically decrypt the data on the server as potential future work.